# A note on Rainich's Conditions in the Null Case

L. Kannenberg
Physics Department
University of Massachusetts Lowell

**ABSTRACT**
It is shown that bivectors are not necessary to satisfy the Rainich conditions in the null case.

---

Rainich's "already unified" classical field theory of gravitation and electromagnetism [1] is based on Einstein's equation for a gravitational field interacting with an electromagnetic field,

$$R_{ij} - \tfrac{1}{2} g_{ij} R_k{}^k = \kappa T_{ij} = \frac{\kappa}{4\pi}\left(F_{ik} F_j{}^k - \tfrac{1}{4} g_{ij} F^k{}_l F_k{}^l\right), \tag{1}$$

where $R_{ij}$ is the Ricci tensor, $T_{ij}$ the electromagnetic stress tensor and $F_{ij} = -F_{ji}$ the electromagnetic field tensor.

Consistency requires that the two sides of Eq. (1) be compatible. One compatibility condition is satisfied trivially; both sides must be symmetric under interchange of the indices $i$ and $j$. Others are not; thus the stress tensor $T_j{}^j$ is identically traceless, which imposes the first of Rainich's algebraic conditions on the Ricci tensor,

$$R_i{}^i = 0, \tag{2}$$

and consequently

$$R_{ij} = \kappa T_{ij}. \tag{3}$$

From the requirement $T^{00} > 0$, usually expressed as

$$v_i T^{ij} v_j > 0 \quad \text{if} \quad v_i v^i < 0, \tag{4}$$

together with the first condition it follows that

$$v_i R^{ij} v_j > 0 \quad \text{if} \quad v_i v^i < 0. \tag{5}$$

This is Rainich's second algebraic condition. Finally, as a consequence of the not-so-obvious identity

$$T^i{}_k T^k{}_j = \delta^i{}_j \, \rho^2, \tag{6}$$

where

$$\rho^2 = \tfrac{1}{4} T^k{}_l T^l{}_k = (8\pi)^{-2}(E^2 - B^2 + 4\mathbf{E}\cdot\mathbf{B})^2, \tag{7}$$

$\mathbf{E}$ and $\mathbf{B}$ being the standard electric and magnetic fields. Eqs. (3) and (6) together produce Rainich's third algebraic condition

---

[1] G. Y. Rainich, *Trans. Am. Math. Soc.* **27**, 106-136 (1925). Published in a mathematics journal and

$$R^i{}_k R^k{}_j = \kappa^2 \, \delta^i{}_j \, \rho^2 \,. \tag{8}$$

It is clear from Eq. (7) that in general $\rho^2 > 0$ except in the unique "null" case for which the two invariants $\mathbf{E}^2 - \mathbf{B}^2$ and $\mathbf{E} \cdot \mathbf{B}$ vanish simultaneously.

In the general case a spatial rotation and exhaustion of the constraints imposed by Eq. (6) followed by a boost casts $T^{ij}$ into a "canonical" frame in which it is diagonalized. From its four independent eigenvectors Rainich constructed six independent bivectors, e.g. $F^{ij} = k^{[i} h^{j]}$, which taken together define a field that satisfies Rainich's three algebraic conditions. In the null case there is no boost that diagonalizes $T^{ij}$, leaving it in the so-called second canonical form

$$[T]^i{}_j = \begin{bmatrix} -\alpha^2 & 0 & 0 & +\alpha^2 \\ 0 & 0 & 0 & 0 \\ 0 & 0 & 0 & 0 \\ -\alpha^2 & 0 & 0 & \alpha^2 \end{bmatrix} \qquad \rho = 0 \tag{9}$$

This matrix has only three independent eigenvectors, from which only three independent bivectors can be formed. As a consequence Rainich's procedure fails if $\rho = 0$; but he dismissed the null case as unphysical in any event, likening the electromagnetic field to an entire function, zero everywhere if its invariants are zero over any given patch.[2]

Finally, the three algebraic conditions imposed on the Ricci tensor by the stress tensor are complemented by the identity

$$D_i(R^i{}_j - \tfrac{1}{2} \delta^i{}_j R^k{}_k) = 0 \tag{10}$$

the geometric side of Eq. (1) imposes on the stress tensor, viz.

$$D_i T^i{}_j = 0 \,. \tag{11}$$

It is well known that Eq. (11) is satisfied if the field tensor $F^{ij}$ satisfies Maxwell's equations,

$$D_i F^{ij} = 0 \tag{12a}$$

$$D_i F_{jk} + D_j F_{ki} + D_k F_{ij} = 0 \,, \tag{12b}$$

so Rainich's final task was to cast his field $F^{ij}$ into a form that satisfies Eqs. (12) while preserving its compliance with the three algebraic conditions. This he achieved by carrying out a duality transformation on his field,

$$F'^{ij} = F^{ij} \cos\theta - {}^*F^{ij} \sin\theta \tag{13}$$

$${}^*F'^{ij} = F^{ij} \sin\theta + {}^*F^{ij} \sin\theta \,, \tag{14}$$

where ${}^*F^{ij}$ is the Hodge dual of $F^{ij}$, and requiring that the dual field satisfies Maxwell's equations.

---

[2] G. Y. Rainich, op. cit., p. 114.

After Misner revived Rainich's work a number of researchers took the null case seriously[3], perhaps motivated in part by the ubiquity in theoretical physics of electromagnetic plane waves, for which both invariants are known to be zero. The fundamental problem was to find an acceptable non-eigenvector substitute for the fourth eigenvector in Rainich's bivector field construction. Their success in this enterprise is epitomized in Torre's magisterial article.

And yet a question remained: Are bivectors the only possible resolution of the null case? The following exercise demonstrates that the answer is no.

It is easy to show that all the eigenvalues of $T^i{}_j$ are zero, so if $u^i$ is any eigenvector of $T^i{}_j$, then

$$T^i{}_j u^j = 0, \tag{15}$$

from which it follows that

$$D_i(T^i{}_j u^j) = (D_i T^i{}_j) u^j + T^i{}_j D_i u^j = 0 . \tag{16}$$

Imposing Eq. (10) on Eq. (16) and recalling that $T^{ij}$ is symmetric produces the condition

$$D_i u_j + D_j u_i = 0 \tag{17}$$

on $u_i$; that is, any eigenfunction of $T^i{}_j$ is a Killing vector. From this follows at once

$$D_i u^i = 0 , \tag{18}$$

that is, the $u^i$ are divergenceless. Furthermore, the divergence of Eq. (17) yields

$$D^i D_i u_j + D_i D_j u^i = 0 . \tag{19}$$

Upon applying Eq. (18) and the general rule

$$D_i D_j v^i - D_j D_i v^i = R_{ij} v^i \tag{20}$$

for any vector $v^i$ to Eq. (19) it follows that

$$D^i D_i u_j + R_{ij} u^i = 0 . \tag{21}$$

Finally, applying Eqs. (3) and (12) to Eq. (18) we are left with

$$D^i D_i u^j = 0 , \tag{22}$$

that is, $u^j$ satisfies the wave equation.

Given Eqs. (17) – (22) it is natural to define the field tensor $F_{ij}$ as

$$F_{ij} = D_i u_j - D_j u_i . \tag{23}$$

---

[3] See for example C. G. Torre, *Class. Quantum Grav.* **31**,1-19 (2014) and references therein.

which automatically satisfies Eq. (11b). The final step is to verify that this form of $F_{ij}$ also satisfies Eq. (11a): We start with

$$D^i F_{ij} = D^i D_i u_j - D^i D_j u_i . \tag{24}$$

From Eq. (22) the first term on the right of Eq. (24) vanishes, leaving us with

$$D^i F_{ij} = - D_i D_j u^i$$

$$= - D_j D_i u^i - R_{ij} u^i \tag{25}$$

upon applying Eq. (20). The first term on the right vanishes by Eq. (18), and the second also vanishes as a consequence of Eqs. (3) and (15), validating Eq. (23) as an acceptable form for the electromagnetic field tensor. The exercise can be run backwards to verify that if the vector potential $A_i$ of a null electromagnetic field $F_{ij}$ is a divergenceless solution of the free wave equation, then it is an eigenvector of the stress tensor $T_i{}^j$.

It is perhaps worth noting explicitly that Eq. (20) maintains the invariance of $F_{ij}$ to the gauge transformation

$$u'_I = u_i + \partial_i \lambda , \tag{26}$$

even though Eq. (18) automatically selects the Lorentz gauge for the vector potential $u_i$ itself.

The generic eigenvector $u_i$ is a superposition of the linearly independent eigenvectors of $T_i{}^j$. From Eq. (14) it is easy to see that there are three, one null and the other two spacelike. It is convenient to label them $u^0{}_i$, $u^1{}_i$, $u^2{}_i$, with the multiplication table

$$u^0{}_i u^{0i} = 0 = u^0{}_i u^{1i} = u^0{}_i u^{2i} = u^1{}_i u^{2i} \tag{27a}$$

$$u^1{}_i u^{1i} = 1 = u^2{}_i u^{2i} . \tag{27b}$$

Observe that the multiplication table is preserved under the duality rotation

$$u^{1\prime} = u^1 \cos\theta - u^2 \sin\theta \tag{28a}$$

$$u^{2\prime} = u^1 \sin\theta + u^2 \cos\theta . \tag{28b}$$

In the canonical frame of Eq. (14) their normalized components are

$$u^0{}_i \doteq 2^{-1/2} (1,0,0,-1) \tag{29a}$$

$$u^1{}_i \doteq (0,1,0,0) \tag{29b}$$

$$u^2{}_i \doteq (0,0,1,0) . \tag{29c}$$

The generic eigenvector then takes the form

$$u_i = a_\mu u^\mu{}_i \tag{30}$$

where $\mu = 0, 1, 2$ and the $a_\mu$ are any real numbers.

To relate the eigenvectors to the metric tensor requires the introduction of another vector, $u^3{}_i$ independent of the $u^\mu{}_i$ and obviously not an eigenvector of $T^i{}_j$, to complete a tetrad. Its components in the canonical frame are easily read off from Eqs. (29):

$$u^3{}_i \doteq 2^{-1/2} (1,0,0,1) \tag{31}$$

The multiplication table of Eqs. (27) is then completed with

$$u^3{}_i u^{3i} = 0 = u^3{}_i u^{1i} = u^3{}_i u^{2i} \tag{32a}$$

$$u^3{}_i u^{0i} = -1 \tag{32b}$$

It is convenient to replace the tetrad $u^a{}_i$, where $a = 0, 1, 2, 3$, with a *vierbein* $e^a{}_i$, an orthonormal basis,

$$e^a{}_i e^i{}_b = \delta^a{}_b \tag{33a}$$

$$e^a{}_i e^j{}_a = \delta^j{}_i, \tag{33b}$$

with components

$$e^a{}_i \doteq \delta^a{}_i \tag{34}$$

in the canonical frame. In terms of this vierbein Eqs. (29) and (31) read

$$u^0{}_i \doteq 2^{-1/2} (e^0{}_i - e^3{}_i) \tag{35a}$$

$$u^1{}_i \doteq e^1{}_i \tag{35b}$$

$$u^2{}_i \doteq e^3{}_i \tag{35c}$$

$$u^3{}_i \doteq 2^{-1/2} (e^0{}_i + e^3{}_i) \tag{35d}$$

which also hold in general. Inverting Eqs. (35) is straightforward:

$$e^0{}_i = 2^{-1/2} (u^0{}_i + u^3{}_i) \tag{36a}$$

$$e^1{}_i = u^1{}_i \tag{36b}$$

$$e^2{}_i = u^2{}_i \tag{36c}$$

$$e^3{}_i = 2^{-1/2} (u^0{}_i - u^3{}_i). \tag{36d}$$

The metric tensor can then be expressed in the familiar form

$$g_{ij} = e^a{}_i \eta_{ab} e^b{}_j. \tag{37}$$

At least two open questions remain. First, what is the relationship between the solution space covered by Torre's bivector procedure and that of the one presented here? And second, for the general case is there a non-bivector procedure analogous to the one presented here? Their answers should be instructive.